\documentclass[12pt]{article}
\usepackage{amsfonts}
\usepackage{amssymb}
\usepackage{amsmath}
\usepackage{amsthm}
\usepackage{graphicx}

\begin{document}

\title{\bf Algebraic aspects of massive gravity}

\author{Alexey Golovnev${}^{a,b}$,  Fedor Smirnov${}^{a}$\\
${}^{a}${\it Faculty of Physics, St. Petersburg State University,}\\ 
{\it Ulyanovskaya ul., d. 1, Saint Petersburg 198504, Russia}\\
{\small agolovnev@yandex.ru} \qquad {\small  sigmar40k@yandex.ru}\\
${}^{b}${\it Asia Pacific Center for Theoretical Physics,}\\ 
{\it Pohang 37673, Republic of Korea}\\
{\small alexey.golovnev@apctp.org}}
\date{}

\maketitle

\begin{abstract}
We describe the freedom of choosing a square root in massive gravity from algebraic point of view. This contribution is based on the talk given by one of us (AG) at the Geometric Foundations of Gravity conference in Tartu in August 2017.
\end{abstract}

\section{Introduction}	

As is well known \cite{Kurt, Claudia, Angnis}, ghost-free models of massive gravity were recently found which have strictly five (or seven in bimetric case) gravitational  degrees of freedom, at least at the classical level. By now, there is an extensive literature on this topic which touches upon  a wide list of directions, from quantum corrections to cosmological solutions. We refer the reader to the review papers cited above. Our concern in this contribution would be about one particular formal aspect of the theory, i.e. the issue of square roots of matrices.

The action of dRGT massive gravity reads \cite{HR}
\begin{equation}
\label{action}
S=\int d^nx\sqrt{-g}\left(R+m^2\sum_{i=0}^n \beta_i e_i (\sqrt{g^{-1}\eta})\right)
\end{equation}
where $R$ is the scalar curvature in Riemannian geometry with metric $g_{\mu\nu}$, parameters $\beta_i$ are arbitrary numbers, and $e_i (\sqrt{g^{-1}\eta})$ are the elementary symmetric polynomials of the eigenvalues $$e_i\equiv\sum\limits_{k_1<k_2<\ldots<k_i}\lambda_{k_1}\lambda_{k_2}\cdots\lambda_{k_i}$$ of a matrix $\Phi^{\mu}_{\nu}$ which is defined by algebraic relation $\Phi^{\mu}_{\alpha}\Phi^{\alpha}_{\nu}=g^{\mu\alpha}\eta_{\alpha\nu}$.

Note that we have dispalyed the simplest case in which the Minkowski metric $\eta$ serves as a fiducial one. One can actually take any fiducial metric $f_{\mu\nu}$ instead of $\eta$ making use of  $e_i (\sqrt{g^{-1}f})$. And moreover, the second metric can be made dynamical by adding its own Einstein-Hilbert term, possibly with another independent Planck mass \cite{biHR}.

Note also that values of $i$ from $1$ to $n-1$ give various BD-ghost-free potentials for the graviton. By definition we assume $e_0=1$ so that the $\beta_0$ term is nothing but a cosmological constant, and analogously $\sqrt{-g}\cdot e_n (\sqrt{g^{-1}f})=\sqrt{-f}$ is either a mere constant in massive gravity, or a cosmological constant term for $f_{\mu\nu}$ in bimetric case.

Of course, there are complicated mathematical problems related to square roots of matrices. And, in particular, one can even find continuously many square roots of a unit matrix. Should we pay any attention to those unusual roots? We think that the answer is in positive. First, we can often be interested in much more complicated cases than the weak gravity around the trivial background with $g^{-1}\eta=\mathbb I$  when the common intuition substantiates the simplest choice of $\sqrt{\mathbb I}=\mathbb I$. Second, the standard proof \cite{proofHR} of ghost freeedom apparently allows for at least some freedom of choosing the square root. And one can even use the auxiliary $\Phi$ field for the Hamiltonian analysis \cite{AG1} which allows to detect at least the primary constraint in the spatial sector.

\section{The issue of square roots}

Square root extraction can better be discussed in complex terms since the canonical representation, modulo similarity transformations, acquires the simple form of Jordan block decomposition, and any non-degenerate matrix does  have a square root when considered over the field of complex numbers. 

There is always a discrete freedom of choosing the sign of $\sqrt{\lambda}$ for each eigenvalue $\lambda$ of the initial matrix. In special situations, when some eigenvalues are equal, one can choose unequal square roots $\sqrt{\lambda_i}\neq\sqrt{\lambda_j}$ for equal eigenvalues $\lambda_i=\lambda_j$. If such is the case, there is a continuous infinite freedom of choosing the square root. Indeed, any non-trivial similarity transformation ${\mathcal M}\to{\mathcal S}^{-1}{\mathcal M}{\mathcal S}$ in the subspace spanned by the i-th and j-th eigenvectors would change the square root matrix while keeping the initial matrix intact.

The latter case presents a big issue for the foundations of massive gravity since the square root matrix ceases to be a smooth function of the initial matrix in any reasonable sense: a generic infinitesimal perturbation reduces the continuous freedom to a finite one. The reason can easily be understood. Taking such square root requires to choose two complementary subspaces, with $+\sqrt{\lambda}$ and $-\sqrt{\lambda}$. When a perturbation separates two eigenvalues, some fixed eigenvectors appear which can be arbitrarily inconsistent with the chosen subspaces.

\subsection{Reality conditions}

Note that a real matrix can have complex eigenvalues with non-vanishing imaginary parts, though in complex-conjugate pairs with Jordan blocks of equal size. In this case the discrete freedom is reduced if considered over the field of real numbers since the square roots of eigenvalues should also be chosen in complex conjugate pairs for the square root matrix to be real.

Negative real eigenvalues are worth of special remarks. In this case the square roots $\sqrt{\lambda}$ are imaginary. The square root matrix might still be chosen real if its eigenvalues could be organised in complex-conjugate pairs. Obviously, this is possible if negative eigenvalues go in pairs. In other words, if (and only if) there is an even number of any kind of Jordan blocks with negative eigenvalues, then a real square root does exist. For example, the $-\mathbb I$ matrix does have a real square root in even dimensions (e.g. symplectic structures), and does not have it in odd dimensions.

Above we have assumed that zero is an even number, so that matrices without negative eigenvalues do obviously have real square roots. The real meaning of the statement is that negative eigenvalues do not preclude existence of real square roots if they go in pairs. However, it is easy to see that those roots are precisely the cases when continuous freedom is present over complex numbers. It gets somewhat reduced under reality condition, and this remaining freedom is further reduced to zero (instead of a finite number) if a perturbation separates the two eigenvalues keeping them real.

For example, the $-{\mathbb I}=\left(\begin{matrix} -1 & 0 \\ 0 & -1 \end{matrix}\right)$ matrix in two dimensions has a whole manifold of square roots with eigenvalues $\pm i$ given by ${\mathcal S}^{-1}\left(\begin{matrix} i & 0 \\ 0 & -i \end{matrix}\right){\mathcal S}$ with any ${\mathcal S}\in GL(2,\mathbb C)$. One of those roots is represented by the standard symplectic matrix  $\left(\begin{matrix} 0 & 1 \\ -1 & 0 \end{matrix}\right)$, and generic real solution reads as ${\mathcal S}^{-1}\left(\begin{matrix} 0 & 1 \\ -1 & 0 \end{matrix}\right){\mathcal S}$ with ${\mathcal S}\in GL(2,\mathbb R)$. However, a perturbation of the form $\left(\begin{matrix} -1+\epsilon & 0 \\ 0 & -1+\delta \end{matrix}\right)$ (or anything else in its similarity class) with small real numbers $\epsilon\neq\delta$ does not admit a real square root. Over complex numbers it admits four roots $\left(\begin{matrix} \pm i\cdot\sqrt{1-\epsilon} & 0 \\ 0 & \pm i\cdot\sqrt{1-\delta} \end{matrix}\right)$ none of which is real.

\subsection{Remarks on degenerate matrices}

Degenerate matrices are special. Such matrix might lack a square root even over the field of complex numbers. Indeed, if a matrix has degenerate Jordan blocks of dimension greater than one, its rank gets reduced by squaring since such Jordan blocks split into degenerate blocks of smaller sizes. For example, the square of the two-dimensional Jordan block with $\lambda=0$ is zero. Therefore, a degenerate matrix which consists of one Jordan block of maximal size does not have any square root at all. More generally, there would be non-trivial combinatorial conditions on the canonical stucture for a square root to exist.

In principle, by the very interpretation, the two metrics must be non-degenerate, and so too the $g^{-1}f$ matrix. However, in Ref.  \cite{GHW} a situation was considered when the fiducial metric dynamically becomes degenerate which has been dubbed a determinant singularity. Given above considerations, we see that the equations could have been evolved through the singularity only due to the highly symmetric nature of the ansatz. In presence of non-trivial Jordan blocks the matrix might cease to have any square root.

\section{Reality and other physical requirements}

Keeping the square root matrix real can be viewed as both too strong and too weak a requirement. On one hand, with a purely imaginary matrix one can have a $\beta_2$ model with real action because it is quadratic in eigenvalues. On the other hand, having real action is probably not the only physical requirement one can think of. We should rather be interested in having a reasonable causal structure of the model. And it should always be possible to consider perturbations around a given solution.

The physical and geometrical meaning of conditions for real square roots is thoroughly discussed in the Ref. \cite{Kocic}. We will only present a few simple observations.

\subsection{Convex cones of Lorentzian structures}

Even if negative eigenvalues go in pairs so that a real root exists, it readily fails under any (infinitesimal) perturbation which separates the negative eigenvalues. Let us note that if $g^{-1}f$ does have a negative eigenvalue $\lambda=-c$ with $c>0$, then the corresponding eigenvector lies in the kernel of $c \cdot g_{\mu\nu}+f_{\mu\nu}$ matrix which gives a degenerate linear combination of $g_{\mu\nu}$ and $f_{\mu\nu}$ with
positive coefficients. 

In other words, $g^{-1}f$ does not have a negative real eigenvalue if and only if the convex cone of $g$ and $f$ in the vector space of matrices over the ordered field of real numbers does not intersect with the hypersurface of degenerate matrices. If we want every element of the convex cone to define a Lorentzian structure, then negative eigenvalues are prohibited.

\subsection{Embedded light cones}

We see that causal properties are tightly related with the algebraic structure of the matrix.
As a clear overshoot, let us require that one of the metrics has its light cone fully inside the light cone of another.  We wish to prove in this case that the matrix  $(g^{-1}f)^{\mu}_{\phantom{..} \nu}$ has a timelike (with respect to both metrics) eigenvector with positive real eigenvalue.

Assume that the light cone of $f$ lies inside that of $g$. Let us also take a $g$-spacelike hypersurface of ${\mathcal F}_1(x)\equiv g_{\mu \nu}x^{\mu}x^{\nu}=const$ in the tangent space, see it as a hyperbolic line on the Figure:

 \includegraphics[width=7cm]{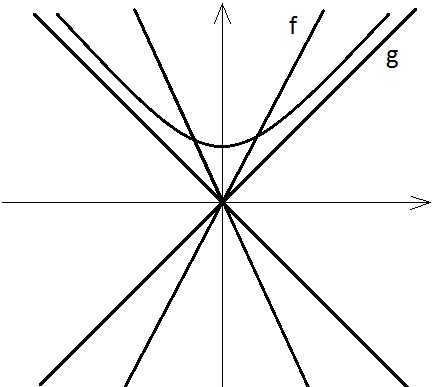}

Consider another function, namely ${\mathcal F}_2(x)\equiv f_{\mu \nu}x^{\mu}x^{\nu}$. It changes its sign while crossing the light cone of $f$. Therefore, assuming everything is smooth and bounded, the function ${\mathcal F}_2(x)$ restricted to the hypersurface ${\mathcal F}_1(x)=const$ reaches an extremal value somewhere inside its intersection with the light cone of $f$. In other words, there is a point $x_{*}$ on the surface at which the gradient $\partial_{\mu}{\mathcal F}_2$ is normal to ${\mathcal F}_1(x)=const$. And therefore $\partial_{\mu}{\mathcal F}_2(x_{*})=C\cdot\partial_{\mu}{\mathcal F}_1(x_{*})$ with some  $C>0$.

The latter condition can be written explicitly as $2 C g_{\mu \nu} x^{\nu}_{*}= 2 f_{\mu \nu} x^{\nu}_{*}$, and implies a timelike eigenvector for $g^{-1}f$ since
$g^{\alpha \mu} f_{\mu \nu} x^{\nu}_{*} = C x^{\alpha}_{*}.$

\subsection{Matrices and eigenvalues}

There is an interesting problem with continuous freedom which has been pointed out in Ref. \cite{Kocic}. If we consider a coordinate transformation $x\to x^{\prime}$, it is easy to see that the matrix $g^{-1}f$ undergoes a similarity transformation.  Therefore, similarity transfomations for those matrices can be interpreted as gauge transformations in the model, and the roots with continuous freedom can be discarded as gauge symmetry violating \cite{Kocic}.

However, if one looks closer at the action principle, the only thing we need are the eigenvalues, not the matrices themselves. And the eigenvalues are not sensitive to similarity transformations. Therefore, if one makes up a formulation of massive gravity which deals directly with eigenvalues without components of the square root matrix, then the argument from Ref. \cite{Kocic} is rebutted and all square roots are equally legitimate from the viewpoint of gauge invariance.

Admittedly, any formulation which requires precise knowledge of eigenvalues seems largely impractical. However, working with elementary symmetric polynomials might be much simpler. We will come back to this point in section 5.

\section{Perturbation theory in massive gravity}

In the standard approach one uses perturbation theory in terms of matrices. In the simplest set-up of weak gravity we have $g^{-1}f={\mathbb I}-{\mathcal H}$ where $\mathcal H$ is a small perturbation matrix. With the trivial choice of $\sqrt{\mathbb I}=\mathbb I$, the usual Taylor expansion works well giving a nicely convergent series 
\begin{equation}
\label{Taylor}
\sqrt{{\mathbb I} - {\mathcal H}}={\mathbb I} -\frac12 {\mathcal H} - \frac18 {\mathcal H}^2 + {\mathcal O}({\mathcal H}^3).
\end{equation}

The working of equation (\ref{Taylor}) hinges upon commutativity of perturbation with the background value of the square root. In less trivial situations, the task becomes more complicated. In particular, if
 $g^{-1}\eta={\mathcal M}^2$ for some choice of the square root $\mathcal M$ at the level of background solution then for linear perturbation we have an equation
\begin{equation}
\label{Sylv}
{\delta\mathcal M}\cdot{\mathcal M}+{\mathcal M}\cdot{\delta\mathcal M}=\delta(g^{-1}\eta)
\end{equation}
of Sylvester type. It works well if and only if the spectra of ${\mathcal M}$ and $-{\mathcal M}$ have empty intersection \cite{Angi}: precisely the case of absence of continuous freedom.

Otherwise the standard perturbation theory is totally ill defined.

\section{Reformulation of massive gravity}

As we have seen, we need to reformulate the model for better incorporation of the mathematical structures it has behind. On the other hand, phenomenological problems such as the absence of reasonable cosmologies \cite{nocosmo} in the simplest set-ups, also call for better study of non-trivial square roots \cite{funnycosmo}.

As we have already discussed, a natural idea would be to formulate the massive gravity action (\ref{action}) directly in terms of eigenvalues. Since eigenvalues are too complicated to find, we have proposed \cite{AGFS1} to work in terms of elementary symmetric polynomials.

The main problem is to relate the polynomials of a matrix $\mathcal M$ such that ${\mathcal M}^2=g^{-1}\eta$ with those of $g^{-1}\eta$. The crucial fact is that the elementary symmetric polynimials are coefficients of the characteristic polynomial:
\begin{equation}
{\rm det}\left({\mathcal M}-\lambda{\mathbb I}\right)=\prod\limits_{i=1}^n\left(\lambda_i-\lambda\right)=\sum_{i=0}^n (-\lambda)^{n-i}\cdot e_i(\mathcal M).
\end{equation}

Then a simple observation
\begin{equation*}
{\rm det}\left({\mathcal M^2}-\lambda^2{\mathbb I}\right)
={\rm det}\left({\mathcal M}-\lambda{\mathbb I}\right)\cdot{\rm det}\left({\mathcal M}+\lambda{\mathbb I}\right)
\end{equation*}
results in the desired relations
\begin{equation}
\label{relation}
\sum\limits_{i+j=2k}(-1)^i e_i(\mathcal M) e_j(\mathcal M)=(-1)^k e_k(\mathcal M^2).
\end{equation}
In four dimensions the explicit form of equations (\ref{relation}) is
\begin{eqnarray*}
e_1(\mathcal M^2) & = & e_1^2(\mathcal M)-2e_2(\mathcal M),\\
e_2(\mathcal M^2) & = & e_2^2(\mathcal M)-2e_1(\mathcal M)e_3(\mathcal M)+2e_4(\mathcal M),\\
e_3(\mathcal M^2) & = & e_3^2(\mathcal M)-2e_2(\mathcal M)e_4(\mathcal M),\\
e_4(\mathcal M^2) & = & e_4^2(\mathcal M).
\end{eqnarray*}

In our formulation \cite{AGFS1} the action (\ref{action}) is interpreted as
\begin{equation*}
S=\int d^nx\sqrt{-g}\left(R+m^2\sum_{i=0}^n \beta_i {\mathfrak e}_i \right)
\end{equation*}
with ${\mathfrak e}_i$ being defined as solutions of the system (\ref{relation}) of quadratic equations:
\begin{equation*}
\sum\limits_{i+j=2k}(-1)^i {\mathfrak e}_i {\mathfrak e}_j=(-1)^k e_k(g^{-1}\eta).
\end{equation*}

The non-uniqueness of the square root is now substituted by having several solutions of equations (\ref{relation}). In particular, one can easily find all solutions \cite{AGFS1} for the square root of the unit matrix which are classified by how many eigenvalues of $\sqrt{\mathbb I}$ are equal to $-1$ instead of $1$. The continuous freedom is luckily ignored by eigenvalues.

It is also clear from this construction that unusual roots break neither gauge invariance nor global symmetries of a background spacetime (Lorentz, rotations, etc.).

\subsection{Perturbation theory in the new approach}

If the Jacobian of $e_i({\mathcal M})\longrightarrow e_i({\mathcal M}^2)$ transformation in formula (\ref{relation}) is non-degenerate at the background solution, then perturbations of $\mathfrak e$ are easily found order by order in perturbation theory. This is the case without continuous freedom, and in particular one can easily reproduce the Fierz-Pauli limit of weak gravity \cite{AGFS1} (if coefficients $\beta_i$ are such that $g_{\mu\nu}=\eta_{\mu\nu}$ is a solution). 

And even if perturbation does not commute with the background square root, our method can be used very straightforwardly when only discrete freedom is present. Each new level of perturbation theory requires solving a simple system of linear equations for corrections to $\mathfrak e$.

For problematic roots the procedure is more complicated due to degeneracy of the Jacobian. Let us assume we have taken $\sqrt{\mathbb I}=\left(\begin{matrix} 1 & 0 \\ 0 & -1 \end{matrix}\right)$ in two dimensions. It gives ${\mathfrak e}_2=-1$ and ${\mathfrak e}_1=0$ for the background.
 
Equations (\ref{relation}) give ${\mathfrak e}^2_2=e_2(g^{-1}\eta)$ and ${\mathfrak e}^2_1-2{\mathfrak e}_2=e_1(g^{-1}\eta)$. Beyond the background level of ${\mathfrak e}_1=0$ we would have two solutions
$${\mathfrak e}_1=\pm\sqrt{e_1(g^{-1}\eta)-2\sqrt{e_2(g^{-1}\eta)}}$$
where we have chosen the sign of ${\mathfrak e}_2$ according to the background value.

We have two branches of solutions. The reason is very simple. Under generic perturbation, two eigenvalues of $\mathbb I$ become unequal, and the sign of ${\mathfrak e}_1$ depends on which one of them is taken with plus (or minus) sign in $\sqrt{\mathbb I}$.

In three dimensions there are three possible choices of one opposite sign eigenvalue which gives three branches as we have explicitly shown in Ref. \cite{AGFS2} having derived a cubic equation for the first order correction to ${\mathfrak e}_1$. In four dimensions there are unusual square roots of unit matrix of two different types: with $1+3$ and $2+2$ signatures. The first case has four branches of perturbations (choosing one eigenvalue out of four), and the second has six (two out of four) \cite{AGFS2}.

Therefore, in our method we can successfully construct perturbations around problematic roots. Note however that there are always several branches and  the limit of $\mathcal H\to 0$ is not analytic as is obvious from the explicit solution for ${\mathfrak e}_1$ in two dimensions, see also our paper \cite{AGFS2}.

\subsection{Split Fierz-Pauli structure}

The structure of quadratic action is easily discussed in terms of eigenvalues of the perturbation matrix. Indeed, the Fierz-Pauli mass term $m^2(h_{\mu\nu}h^{\mu\nu}-(h^{\mu}_{\mu})^2)$ is given by nothing but the second elementary symmetric polynomial of the perturbation matrix $e_2(\mathcal H)=\sum\limits_{i<j}\lambda_i \lambda_j$.

For a non-standard $\sqrt{\mathbb I}$ we generically cannot find $\sqrt{{\mathbb I}-{\mathcal H}}$ by Taylor expanding (let alone the fact it is not well defined at all). However, there is a class of perturbations for which it is possible. These are those which commute with $\sqrt{\mathbb I}$. Let us better understand their structure.

In any square root of $\mathbb I$, we choose two complementary subspaces such that $\sqrt{\mathbb I}$ can be represented in a suitable basis as $\left(\begin{matrix} \mathbb I & \mathbb O \\ \mathbb O & -\mathbb I \end{matrix}\right)$, or $\sqrt{\mathbb I}={\mathbb I}\oplus (-\mathbb I)$. If the perturbation matix is block diagonal in the same basis, ${\mathcal H}={\mathcal H}_1\oplus{\mathcal H}_2$, then $\sqrt{\mathbb I}$ and $\mathcal H$ do commute, and the Taylor expansion works well.

If such is the case, the computation goes the standard way in two independent blocks, and the quadratic action for weak gravity becomes $$C_1\cdot e_2({\mathcal H}_1)+C_2\cdot e_2({\mathcal H}_2)$$ which is the sum of two Fierz-Pauli blocks.

In terms of the full  matrix $\mathcal H$, we have separation of its eigenvalues into two groups and the quadratic action of the form
$$C_1\cdot\sum\limits_{i<j;\ i,j\in {\mathcal I}_{-}}\lambda_i\lambda_j+C_2\cdot\sum\limits_{i<j;\ i,j\in {\mathcal I}_{+}}\lambda_i\lambda_j.$$ 
In our method, this result is actually independent of validity of Taylor expansion. Indeed, the right hand side of equations (\ref{relation}) for ${\mathcal M}^2=\mathbb I - \mathcal H$ depends only on eigenvalues of the matrix $\mathcal H$, and not on directions of its eigenvectors.

Let us also formulate it another way around. The Taylor expansion with a given perturbation $\mathcal H$ works well only if eigenspaces  of $\sqrt{\mathbb I}$ were chosen appropriately. If such is the case, the eigenvalues of $\sqrt{\mathbb I - \mathcal H}$ coincide with those algebraically found from equations (\ref{relation}) which always gives a well defined solution. We have explicitly checked this result \cite{AGFS2} by direct calculations in dimensions three and four.

\section{Discussion and Conclusions}

It remains to be seen whether unusual square roots could be phenomenologically interesting, see also \cite{funnycosmo}. But it is obvious that they are a very important part of fundamental understanding of massive gravity. As we have shown, they are not related with bad pathologies such as gauge symmetry breaking, contrary to another claim in Ref. \cite{Kocic}, however it is not totally obvious that existing proofs of BD-ghost freedom apply to cases when the language of matrices breaks down.

We conclude that the issue of unusual square roots should further be studied. From the non-perturbative point of view, it is unnatural to discard them right away. In Ref. \cite{Kocic} it was suggested to take one branch of the square root function for all eigenvalues. This recipe might not even seem to be well-defined. It depends on the chosen branch cut for the square root function. Generically, it is nothing but a common habit to have it along the negative real half-axis. In principle, the position of the branch cut is a mere immaterial convention. The Riemann surface is everywhere smooth except for the branching point. However, in the context of massive gravity, we know that negative eigenvalues of $g^{-1}\eta$ are problematic, and therefore this choice of the branch cut gets justified.

When a pair of eigenvalues of  $g^{-1}\eta$ crosses the negative real half-axis, perturbations are not well defined even in terms of the eigenvalues (if the reality condition for elementary symmetric polynomials is imposed). Thus, the different branches of $\sqrt{\lambda}$ should probably be considered as different theories\footnote{AG is grateful to Valery Rubakov for raising a question of this type.}.

If a pair of eigenvalues could be dynamically driven towards crossing the branch cut, it would mark a breakdown of the model and require embedding massive (bimetric) gravity into a somewhat more fundamental framework. Probably, there could be some quantum protection mechanism similar to possible protection \cite{causyes} against causal pathologies \cite{causno}. 

We conclude that many foundational issues of massive gravity do require further investigation.

\section*{Acknowledgments}

AG is grateful to the organisers and participants of the Geometric Foundations of Gravity 2017 conference in Tartu for the very fruitful discussions and exciting atmosphere, and to Saint Petersburg State University for travel grant 11.41.1057.2017.


\begin{thebibliography}{99}
\bibitem{Kurt} K. Hinterbichler, Theoretical aspects of massive gravity, {\it  Rev.  Mod. Phys.} {\bf  84} (2012), 671-710.

\bibitem{Claudia} C. de Rham, Massive gravity, {\it Living Reviews in Relativity} {\bf  17} (2014),
7.

\bibitem{Angnis} A. Schmidt-May, M. von Strauss, Recent developments in bimetric
theory, {\it J. Phys. A}
{\bf 49} (2016), 183001.

\bibitem{HR} S.F. Hassan, R.A. Rosen, On non-linear actions for massive gravity,
{\it J. High Energy Phys.} (2011), JHEP07(2011)009.

\bibitem{biHR} S.F. Hassan, R.A. Rosen, Bimetric gravity from ghost-free massive
gravity, {\it J. High Energy Phys.} (2012), JHEP02(2012)126.

\bibitem{proofHR} S.F. Hassan, R.A. Rosen, resolving the ghost problem in non-Linear
massive gravity, {\it Phys. Rev. Lett.} {\bf 108} (2012), 041101.

\bibitem{AG1} A. Golovnev, On the Hamiltonian analysis of non-linear massive
gravity, {\it Phys. Lett. B} {\bf 707} (2012), 404-408.

\bibitem{GHW} P. Gratia, W. Hu, M. Wyman, Self-accelerating massive gravity: how zweibeins walk through determinant singularities, {\it Class. Quantum Grav.} {\bf 30} (2013), 184007.

\bibitem{Kocic} S.F. Hassan, M. Kocic, On the local structure of spacetime in ghost-free bimetric theory and massive gravity, arxiv.org/abs/1706.07806.

\bibitem{Angi} L. Bernard, C. Deffayet, A. Schmidt-May, M. von Strauss, Linear spin-2 fields in most general backgrounds, {\it  Phys. Rev. D} {\bf  93} (2016), 084020.

\bibitem{nocosmo} G. D'Amico, C. de Rham, S. Dubovsky, G. Gabadadze, D. Pirtskhalava, A.J. Tolley, Massive cosmologies, {\it Phys. Rev. D} {\bf 84} (2011), 124046.

\bibitem{funnycosmo} D. Comelli, M. Crisostomi, K. Koyama, L. Pilo, G. Tasinato, New branches of massive gravity,  {\it Phys. Rev. D} {\bf 91} (2015), 121502.

\bibitem{AGFS1} A. Golovnev, F. Smirnov, Dealing with ghost-free massive gravity without explicit square roots of matrices, {\it Phys. Lett. B} {\bf 770} (2017), 209 - 212.

\bibitem{AGFS2} A. Golovnev, F. Smirnov, Unusual square roots in the ghost-free theory of massive gravity,  {\it J. High Energy Phys.} (2017), JHEP06(2017)130.

\bibitem{causyes} C. Burrage, C. de Rham, L. Heisenberg, A.J. Tolley, Chronology Protection in Galileon Models and Massive Gravity, {\it J. Cosm. Astropart. Phys} (2012), JCAP07(2012)004

\bibitem{causno} S. Deser, K. Izumi, Y.C. Ong, A. Waldron, Problems of massive gravities, {\it Mod. Phys. Lett. A} {\bf 30} (2015), 1540006.

\end{thebibliography}
\end{document}